# Automated Multi-Channel Segmentation for the 4D Myocardial Velocity Mapping Cardiac MR

Yinzhe Wu [a,b], Suzan Hatipoglu [c], Diego Alonso-Álvarez [d], Peter Gatehouse [a,c], David Firmin [a,c], Jennifer Keegan [a,c], Guang Yang [a,c]

[a] National Heart & Lung Institute, Imperial College London, SW7 2AZ, London, UK
[b] Department of Bioengineering, Faculty of Engineering, Imperial College London, SW7 2AZ, London, UK
[c] Cardiovascular Biomedical Research Unit, Royal Brompton Hospital, SW3 6NP, London, UK
[d] Research Computing Service, Information & Communication Technologies, Imperial College London, SW7 2AZ, London, UK

**ABSTRACT**

Four-dimensional (4D) left ventricular myocardial velocity mapping (MVM) is a cardiac magnetic resonance (CMR) technique that allows assessment of cardiac motion in three orthogonal directions. Accurate and reproducible delineation of the myocardium is crucial for accurate analysis of peak systolic and diastolic myocardial velocities. In addition to the conventionally available magnitude CMR data, 4D MVM also acquires three velocity-encoded phase datasets which are used to generate velocity maps. These can be used to facilitate and improve myocardial delineation. Based on the success of deep learning in medical image processing, we propose a novel automated framework that improves the standard U-Net based methods on these CMR multi-channel data (magnitude and phase) by cross-channel fusion with attention module and shape information based post-processing to achieve accurate delineation of both epicardium and endocardium contours. To evaluate the results, we employ the widely used Dice scores and the quantification of myocardial longitudinal peak velocities. Our proposed network trained with multi-channel data shows enhanced performance compared to standard U-Net based networks trained with single-channel data. Based on the results, our method provides compelling evidence for the design and application for the multi-channel image analysis of the 4D MVM CMR data.

**Keywords:** Cardiac MRI, Cardiac image segmentation, magnetic resonance imaging (MRI), Multichannel, Machine Learning, Computer-Aided Diagnosis, Medical Imaging Analysis, Image Processing

## 1. INTRODUCTION

Four-dimensional (4D) myocardial velocity mapping (MVM) is a promising Cardiac Magnetic Resonance (CMR) technique to assess cardiovascular conditions. In addition to conventional magnitude data, this technique additionally acquires velocity-encoded phase data, which are used to generate velocity maps in three orthogonal directions throughout the cardiac cycle. Hence, "4D MVM" here refers to the acquisition of three directions of velocity (3D) over time (1D).

Delineating the left ventricle (LV) myocardium in the CMR data is crucial in the determination of the myocardial motion from this data. Manual segmentation of both endocardial and epicardial borders of the LV myocardium is a very tedious process taking up to an hour for an experienced clinician to complete a single-slice 50-frame study. Inspired by the recent development of deep Convoluted Neural Networks (CNNs) in medical image processing [1], in particular the U-Net [2], we here assess and prove that the U-Net based model can be applied to 4D MVM CMR single-channel phase images and single-channel magnitude images. Furthermore, motivated by the growing interests of multi-channel processing in biomedical imaging, we advance our method to encode both magnitude and phase data from 4D MVM CMR with the implementation of attention modules. In addition, we propose a model to incorporate dense connections on all layers of depth of CNNs instead of the traditional single-layer fusion to model the relation among all channels. This model is able to load and infer on each cine slice within less than 15 seconds. The traditional strategy assumes simple (e.g., linear) relationships between channels, which may not represent the reality correctly [3].

In evaluating the LV contours (epicardium and endocardium contours) generated against the manually-delineated ground truth (initialized via an active contour segmentation), we employ both the widely used method of Dice scores and our novel comparison methodology using the quantitative myocardium CMR velocity mapping (MVM) [4] data generated. By evaluating model generated segmentation on the basis of the accuracy of the CMR MVM results generated, we can observe and further compare the segmentation performance on clinically important parameters. Despite the remarkable performance of existing methods on LV myocardium segmentation on CMR single-channel magnitude images, the combination of multi-channel data of 4D MVM CMR at various levels of abstraction has not been exploited. Our paper addresses this problem by assessing, investigating and comparing different strategies of incorporating and analyzing CMR multi-channel data using both Dice scores and clinically relevant CMR MVM results.

To summarize, the novelty of this work is of 3-fold: (1) We have assessed and proved that U-Net can be applied to single-channel magnitude data, single-channel phase data and their combination as multi-channel data respectively from 4D MVM CMR for automated LV myocardium segmentation; (2) We have designed and proposed a novel multi-channel attention block (MMAB) and a novel segmentation framework to better incorporate and to segment the multi-channel data from 4D MVM CMR; (3) We have evaluated clinical accuracy of novel segmentation framework by comparing outputs with the myocardial velocities obtained from manual segmentation.

## 2. METHODOLOGY

### 2.1 Data Acquisition and Preprocessing

The CMR data were acquired from 18 healthy subjects (8 of them were acquired twice, giving 26 datasets in total) at the Royal Brompton Hospital. In each subject, high temporal resolution (50 frames per cardiac cycle) cine spiral MVM with non-Cartesian SENSE reconstruction and 3 orthogonal directions of velocity encoding were acquired in a single breath-hold [5]. The acquired spatial resolution was 1.7mm×1.7mm and data were reconstructed on a 512×512 matrix (reconstructed spatial resolution 0.85 mm×0.85 mm). Data were acquired in short-axis slices from base to apex of the left ventricle. The ground truth of the myocardium segmentation was performed by an experienced CMR physicist. We augmented the data by random horizontal flipping (probability of 0.5) and random rotation (angle =[0,90°]) prior to the training.

### 2.2 Network Architectures for the Myocardium Segmentation

We summarized the available standard strategies on supervised learning-based deep learning as below

[a] U-Net with CMR single-channel magnitude data (512×512×1);

[b] U-Net with CMR single-channel phase data (512×512×3);

[c] U-Net with CMR multi-channel data, where we stacked the magnitude (512×512×1) and phase (512×512×3)

data on the channel axis of the input image.

In addition, we designed the strategies below allowing the model to understand the more latent relationships between different channels.

[d] We proposed a novel network structure (Figure 1) based on standard U-Net structures and its encoders and decoders, where we input the magnitude (512×512×1) and phase (512×512×3) data separately into the network. Each input had its respective encoder to allow the model to extract information that would otherwise be fused at early stages. We then added a novel multi-channel attention block (MMAB) to disentangle the multi-channel information better before it entered the decoder following standard U-Net structure at each depth of the network.

Within MMAB, we firstly concatenated the outputs of two encoders at each depth and added a convolutional block that was of the same setting as the encoder to allow the model to disentangle the latent relationship between the information between the two channels. Then, the convoluted data stream went through an attention block [6] before being delivered to the decoder at each depth.

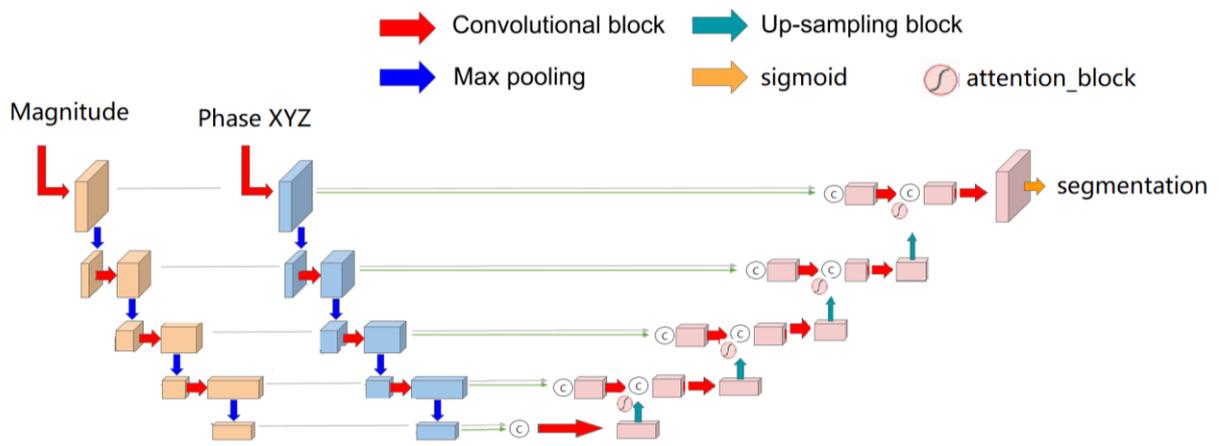

Figure 1. Network architecture of our proposed myocardium segmentation in 4D MVM CMR.

## 2.3 Training Procedure

We stacked all cine CMR images on their frame axis and gives magnitude and phase data, each of 22750 frames. All models specified in Section 2.2 were trained from scratch and evaluated using five-fold cross-validation from the dataset. We trained our network with cross-entropy loss with Adam optimizer. For all experiments, we defined the batch size as 8 to accommodate the high image resolution of 512×512. The model was trained for 5 epochs.

## 2.4 Post-processing

The endocardial and epicardial borders of the CMR short-axis data are approximately ellipsoid. Any 'broken' segmentations, which did not conform to this pass through an additional post-processing stage. Examples of 'broken' segmentation included semi-ellipses and ellipses that were not closed.

At the post-processing stage, all but the largest connected component on each 2D frame were automatically removed on predicted segmentation. Then we followed the mathematical geographical regression formula in cartesian coordinates to regress the predicted segmentation into a 3-pixel width ellipse, and we then output the final prediction as the regressed ellipse on top of the original predicted segmentation. Finally, we took the coordinates of the epicardium and the endocardium layer for the generation of CMR MVM.

**2.5 Evaluation Metrics**

2.5.1 Dice Similarity Coefficient (or Dice scores)

To assess and compare the performance of each model, we first evaluated accuracy of predictions using Dice scores on raw predicted segmentation data before post-processing (1) per cine frame, (2) per 2D slice (50 cine frames), and (3) per subject (3–5 slices). The three Dice score metrics allowed a comprehensive observation and examination of the model's performance.

2.5.2 CMR MVM Data

To assess the clinical significance of the predicted segmentation and its accuracy for obtaining clinical measurements , we compared three characteristic values, i.e., peak systolic (PS), peak diastolic (PD) and peak atrial systolic (PAS) longitudinal velocities of the global estimated longitudinal velocity curves with the same values obtained after manual segmentation.

## 3. EXPERIMENTS AND RESULTS

We cross-validated all four models (Section 2.2 [a]–[d]) to demonstrate their capabilities in CMR MVM myocardium segmentation using the two evaluation methodologies. Then we compared the results of different models by the mean and standard deviation. Results were tested for statistically significant differences using the Wilcoxon signed-rank tests, and the tests were considered significant if associated $p$ values were <0.05.

The quantitative results of the models validated by Dice scores can be found in Table 1. By comparing the results between single-channel input based models in Section 2.2 [a] and [b] and the multi-channel input based models Section 2.2 [c] and [d], we can see that the multi-channel models [c] and [d] yielded higher Dice scores, where model [d] resulted in the highest in particular. The performance of models [c] and [d] is confirmed by the statistical significance ($p<0.05$) when comparing the per cine frame Dice scores of [c] and [d] against the results of [a] and [b] (Figure 2B(c)). Our proposed model [d] yielded the highest mean Dice scores, confirmed by the statistical significance ($p<0.05$) when comparing the per cine frame Dice scores between models [c] and [d] (Figure 2B).

The statistical test results for quantitative results of myocardial longitudinal peak velocities of different models in the framework can be found in Table 2, where we considered the segmentation model and the post-processing stage as one segmentation framework. In evaluation of our segmentation framework, we can observe high correlation coefficients (R) and statistically significantly linear relationships ($p<0.05$) across all models in all parameters (PS, PD and PAS) (Table 2). In particular, we found model [a] had the highest R for PS data, whereas model [c] had the highest R for PD and PAS data. We also conducted Bland-Altman tests for each of the PS, PD, and PAS values generated from difference models, which showed only few outliers out of the range of mean +/- 1.96 standard deviation. The Bland-Altman analysis of the model [d] can be found in Figure 3. The Bland-Altman plots for the other models are similar to the ones shown in this figure.

Table 1. Statistics (mean and standard deviation) Dice scores of different models prior to post-processing.

| Model Type (2.2) / Dice Score Type | Single-Channel Data | | Multi-Channel Data | |
|---|---|---|---|---|
| | [a] (Dice score mean, standard deviation) | [b] (Dice score mean, standard deviation) | [c] (Dice score mean, standard deviation) | [d] (Dice score mean, standard deviation) |
| Per subject | 0.8604, 0.0324 | 0.8513, 0.0292 | 0.8672, 0.0271 | **0.8679, 0.0290** |
| Per cine slice | 0.8615, 0.0687 | 0.8492, 0.0784 | 0.8688, 0.0551 | **0.8692. 0.0489** |
| Per cine frame | 0.8596, 0.0850 | 0.8455, 0.0977 | 0.8666, 0.0713 | **0.8668, 0.0747** |

Table 2. Statistical tests of myocardium CMR velocity mapping (MVM) characteristic values (PS, PD, and PAS per cine slice) between the automated segmented contours and the ground truth.

| Correlation coefficient (R) and *p*-value calculated between the model's and the manual ground truth's CMR MVM PS, PAS and PD values per cine slice | | | | | | | | |
|---|---|---|---|---|---|---|---|---|
| | [a] | | [b] | | [c] | | [d] | |
| | R coefficient | *p*-value | R coefficient | *p*-value | R coefficient | *p*-value | R coefficient | *p*-value |
| **PS** | 0.9239 | <0.0001 | 0.9073 | <0.0001 | 0.8649 | <0.0001 | 0.8870 | <0.0001 |
| **PD** | 0.9885 | <0.0001 | 0.9870 | 0.0026 | 0.9979 | <0.0001 | 0.9864 | <0.0001 |
| **PAS** | 0.9901 | <0.0001 | 0.9905 | <0.0001 | 0.9989 | <0.0001 | 0.9752 | <0.0001 |

## 4. CONCLUSIONS

In this study, we have proved that the standard U-Net based structure can be used for myocardium segmentation in 4D MVM CMR. In addition, we have proposed a multi-channel attention block to better incorporate the multi-channel information and have used that to design our novel network architecture with post-processing stages to enhance the segmentation accuracy. In order to assess the performance of the designed frameworks, we have evaluated them with relation to quantitative CMR MVM velocity peaks in addition to the Dice scores. Our experiment results have demonstrated the high efficacy of our proposed automated segmentation framework.

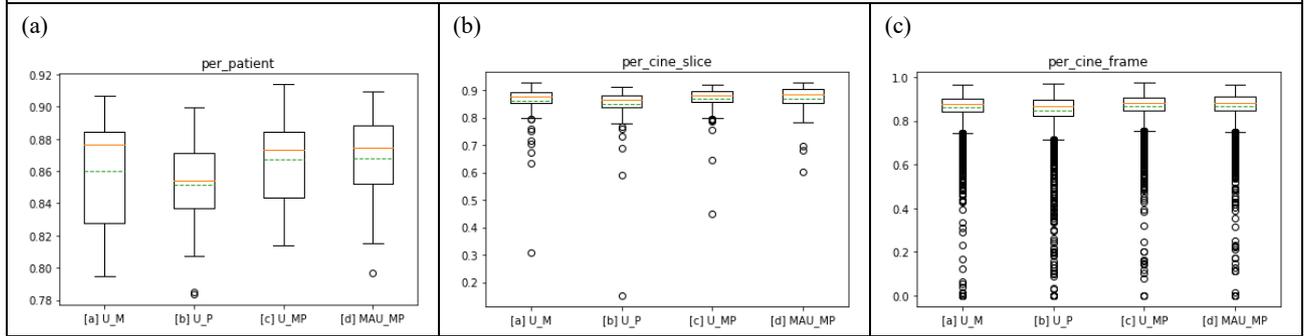

Figure 2. Comparison study results using different methods.

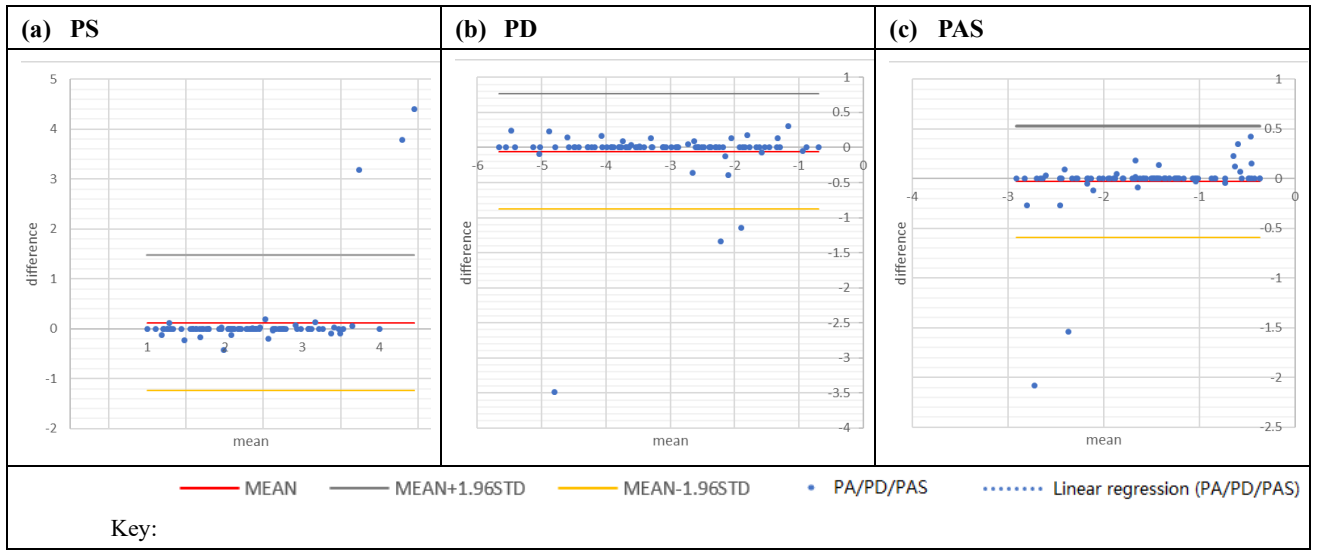

Figure 3. Bland–Altman plot of CMR MV PS, (a), PD, (b), and PAS, (c), values (between data generated from the model's segmentation and the manual segmentation) for model [d] (difference = model − manual).